\documentclass[preprint,aps,prb,showpacs,amsfonts,amssymb,superscriptaddress]{revtex4}
\usepackage{graphicx}
\usepackage{stmaryrd}

\begin{document}
\title{Optical phonon scattering of cavity polaritons in an electroluminescent device}
\author{A. Delteil}
\affiliation{Laboratoire ``Mat\'eriaux et Ph\'enom\`enes
Quantiques'', Universit\'e Paris Diderot, CNRS-UMR 7162, 75013
Paris, France}
\author{A. Vasanelli}
\email{Angela.Vasanelli@univ-paris-diderot.fr}
\affiliation{Laboratoire ``Mat\'eriaux et Ph\'enom\`enes
Quantiques'', Universit\'e Paris Diderot, CNRS-UMR 7162, 75013
Paris, France}
\author{P. Jouy}
\affiliation{Laboratoire ``Mat\'eriaux et Ph\'enom\`enes
Quantiques'', Universit\'e Paris Diderot, CNRS-UMR 7162, 75013
Paris, France}
\author{D. Barate}
\affiliation{Laboratoire ``Mat\'eriaux et Ph\'enom\`enes
Quantiques'', Universit\'e Paris Diderot, CNRS-UMR 7162, 75013
Paris, France}
\author{J.C. Moreno}
\affiliation{Institut d'Electronique du Sud, Universit\'e de Montpellier 2, 34095  Montpellier, France}
\author{R. Teissier}
\affiliation{Institut d'Electronique du Sud, Universit\'e de Montpellier 2, 34095  Montpellier, France}
\author{A.N. Baranov}
\affiliation{Institut d'Electronique du Sud, Universit\'e de Montpellier 2, 34095  Montpellier, France}
\author{C. Sirtori}
\affiliation{Laboratoire ``Mat\'eriaux et Ph\'enom\`enes
Quantiques'', Universit\'e Paris Diderot, CNRS-UMR 7162, 75013
Paris, France}


\begin{abstract}
A signature of the scattering between microcavity polaritons and longitudinal optical phonons has been observed in the electroluminescence spectrum of an intersubband device operating in the light-matter strong coupling regime. By electrical pumping we resonantly populate the upper polariton branch at different energies as a function of the applied bias. The electroluminescent signal arising from these states is seconded by a phonon replica from the lower branch.
\end{abstract}

\maketitle
The strong coupling between light and matter in semiconductor microcavities gives rise to new quasiparticles, the microcavity polaritons~\cite{weisbuch_PRL}. Owing to their bosonic character, these particles can undergo stimulated scattering towards a final state~\cite{savvidis} and give rise to a polariton laser~\cite{christopoulos,bajoni}. One of the mechanisms that have been exploited for stimulated scattering is the optical phonon assisted relaxation, from an optically pumped excitonic reservoir into a minimum of the polaritonic dispersion curve~\cite{imamoglu, boeuf, maragkou}. Recently~\cite{deliberato_PRL}, it has been theoretically shown that optical phonons can be also at the origin of stimulated scattering of intersubband polaritons~\cite{dini}. This scheme constitutes the base for the realisation of novel mid infrared lasers without electronic population inversion. In this work, we present an experimental signature of the scattering between intersubband polaritons and longitudinal optical (LO) phonons in an electrically injected quantum cascade device~\cite{sapienza_PRL}. While electrons are resonantly injected in the upper polariton states only, the electroluminescence (EL) spectrum is composed of two features, from both branches, separated by the energy of an optical phonon. This scattering event shines light on the nature of the intersubband polaritons, in which the matter part interacts efficiently with the ions of the lattice producing a profound variation of the photonic emission.

\begin{figure}[ht]
\includegraphics[width=0.8\columnwidth]{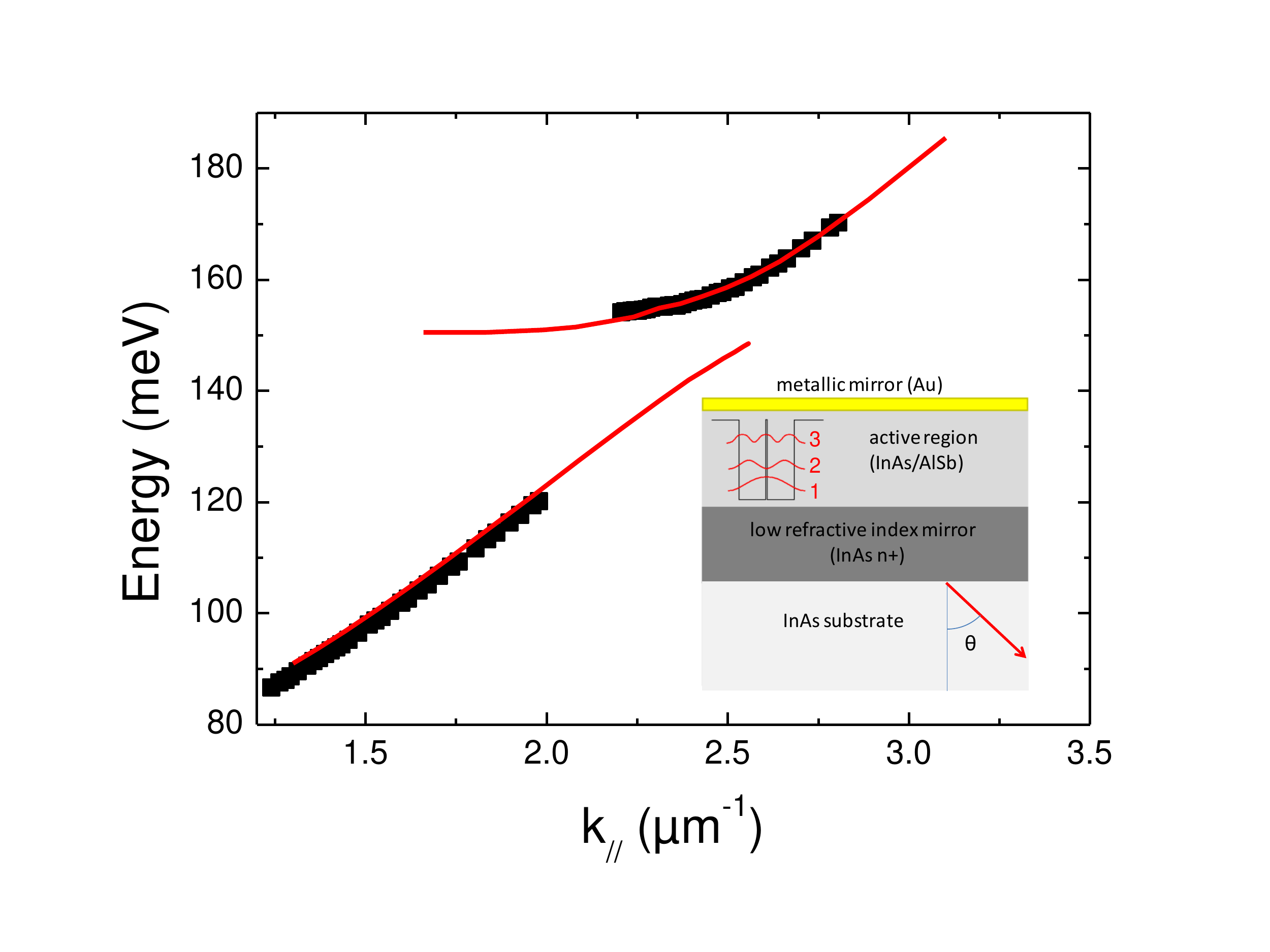}
\caption{(Colors online) Main panel: Measured and simulated polaritonic dispersion. Squares: absorption maxima measured at 77 K. (Red) Continuous line: Absorption maxima simulated by transfer matrix. Inset: Schematic view of the sample. An InAs/AlSb quantum cascade structure is inserted in a planar microcavity, based on a metallic Au mirror and on a low refractive index InAs mirror. The structure is grown on a non-intentionally doped InAs substrate. $\theta$ is the internal angle for light propagation. }
\label{struc}
\end{figure}

Our device is composed of a 20-period InAs/AlSb quantum cascade structure, imbedded into a planar microcavity. The latter consists of a top metallic mirror evaporated on the surface and a bottom reflector made of a doped InAs layer, as shown in the inset of figure~\ref{struc}. Each period of the cascade is based on a main well and an injection/extraction region, strongly doped in order to have a 2D electron gas in the fundamental subband. The device is designed to ensure a light-matter strong coupling regime between the $1 \rightarrow 2$ transition (at energy $E_{21}=151$~meV) and the first order mode of the planar microcavity. Angle resolved multipass transmission experiments at 77~K have been performed to characterize the structure. The energy position of the absorption maxima are plotted in figure~\ref{struc} (squares) with respect to the in-plane photon wave vector $k_{\sslash}$. They show a clear anticrossing between the intersubband transition and the cavity mode, with a vacuum field Rabi splitting $2 \, \hbar \Omega_R=13$~meV. The absorption is simulated by using transfer matrix formalism (line): a very good agreement with the experimental results is obtained by using an electronic density in the fundamental subband $N_s=1.2 \times 10^{12}$~cm$^{-2}$ (the nominal electronic density is $2 \times 10^{12}$~cm$^{-2}$).

\begin{figure}[ht]
\centering
\includegraphics[width=0.8\columnwidth]{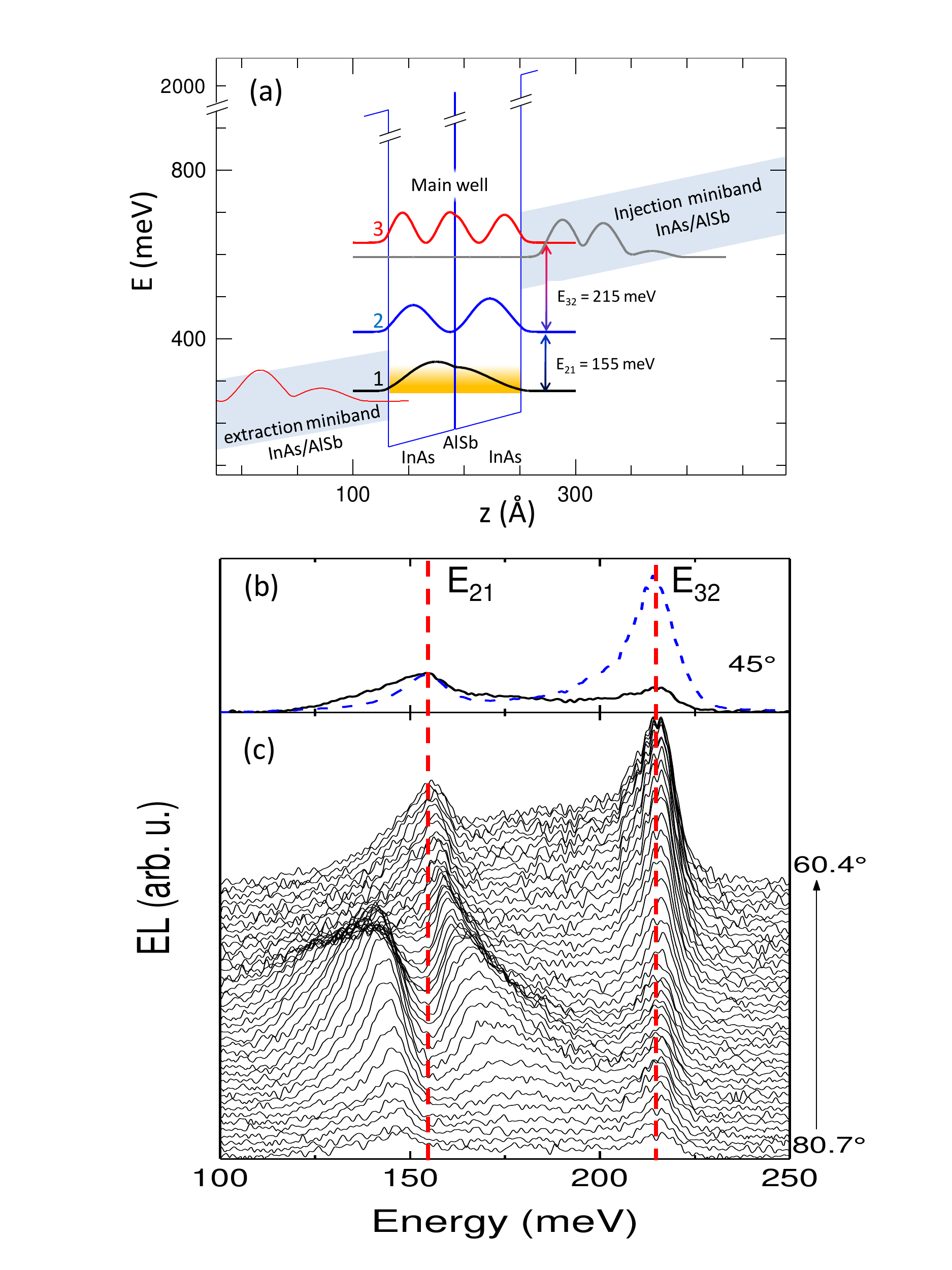}
\caption{(Colors online) (a) Schematic band diagram of one period of the quantum cascade structure under an electric field of 70 kV/cm. The square moduli of the relevant wavefunctions are shown. The layer sequence of one period of the
structure, in nm, starting from the main quantum well, is 5.9/{\textbf{0.1}}/5.9/{\textbf{2.7}}/5.5/{\textbf{0.9}}/5.2/{\textbf{1.2}}/\underline{5.0}/{\textbf{1.2}}/\underline{4.5}/{\textbf{1}}/\underline{3.9}/{\textbf{1.2}}/\underline{3.6}/{\textbf{0.9}}/\underline{3.5}/{\textbf{1.2}}/3.2/{\textbf{0.9}}/3.1/{\textbf{2}}.
AlSb layers are in bold, underlined layers are Te doped 10$^{18}$~cm$^{-3}$.
This sequence has been repeated 20 times in the sample. (b) EL spectra measured at 45$^\circ$ internal angle at 8~V (600 mA) (dashed line) and 6 V (58 mA) (continuous line). (c) Angle-resolved EL spectra measured at 6 V.}
\label{EL}
\end{figure}

In order to perform angle resolved EL spectra, the sample is etched into square mesas of 160~$\mu$m side, mechanically polished, indium soldered onto a copper holder and mounted in a cryostat. All the experimental results shown in this work have been obtained at 77~K. The EL signal from the substrate is collected with a $f/2$ ZnSe lens, analyzed by a Fourier transform infrared spectrometer and detected using a HgCdTe detector through a $f/0.5$ ZnSe lens. A schematized band diagram of the quantum cascade structure under applied bias is shown in figure~\ref{EL}a. When the structure is aligned, electrons in the injector state can tunnel into the subband 3 of the main quantum well. Electrons can then relax radiatively or non-radiatively (mainly by LO-phonon emission) into the subbands 2 and 1 in the main well. The extraction miniband ensures the electronic transport in the following period of the cascade. A very thin AlSb barrier (of nominal thickness 1 {\AA}) is grown in the middle of the main well. The role of this barrier is that of separating in energy the $3 \rightarrow 2$ and $2 \rightarrow 1$ transitions. In fact only wavefunctions corresponding to odd quantum numbers are modified by the presence of this barrier, that increases the energy of the $3 \rightarrow 2$ transition and reduces that of the $2 \rightarrow 1$ transition of the quantum well~\cite{marzin}. The (blue) dashed curve in fig.~\ref{EL}b shows the EL spectrum measured at 8~V (at this voltage the injector is resonant with state 3) at 77~K, at an angle of $45^ \circ$. At this angle the intersubband transition and the cavity mode are far from resonance~\cite{jouy_PRB}, allowing us to observe the bare intersubband transitions. The measured energies, $E_{21}$=154~meV and $E_{32}$=214~meV, are in very good agreement with the ones calculated with a three band $\vec{k} \cdot \vec{p}$ model~\cite{faugeras}, by taking into account collective effects (the simulated transition energies are $E_{21}=151$~meV and $E_{32}=213$~meV). When the bias applied to the structure is reduced, the injector and subband 3 are not aligned anymore, and the intensity of the EL peak at energy $E_{32}$ drops with respect to that at $E_{21}$, as shown in fig.~\ref{EL}b by the (black) continuous curve, measured at 6~V. Note that in this case the EL spectrum presents a wide feature at energies comprised between $E_{21}$ and $E_{32}$. This is related to the diagonal transition between the injector state and the subband 2, which is strongly dependent on the applied voltage.

\begin{figure}[ht]
\centering
\includegraphics[width=0.8\columnwidth]{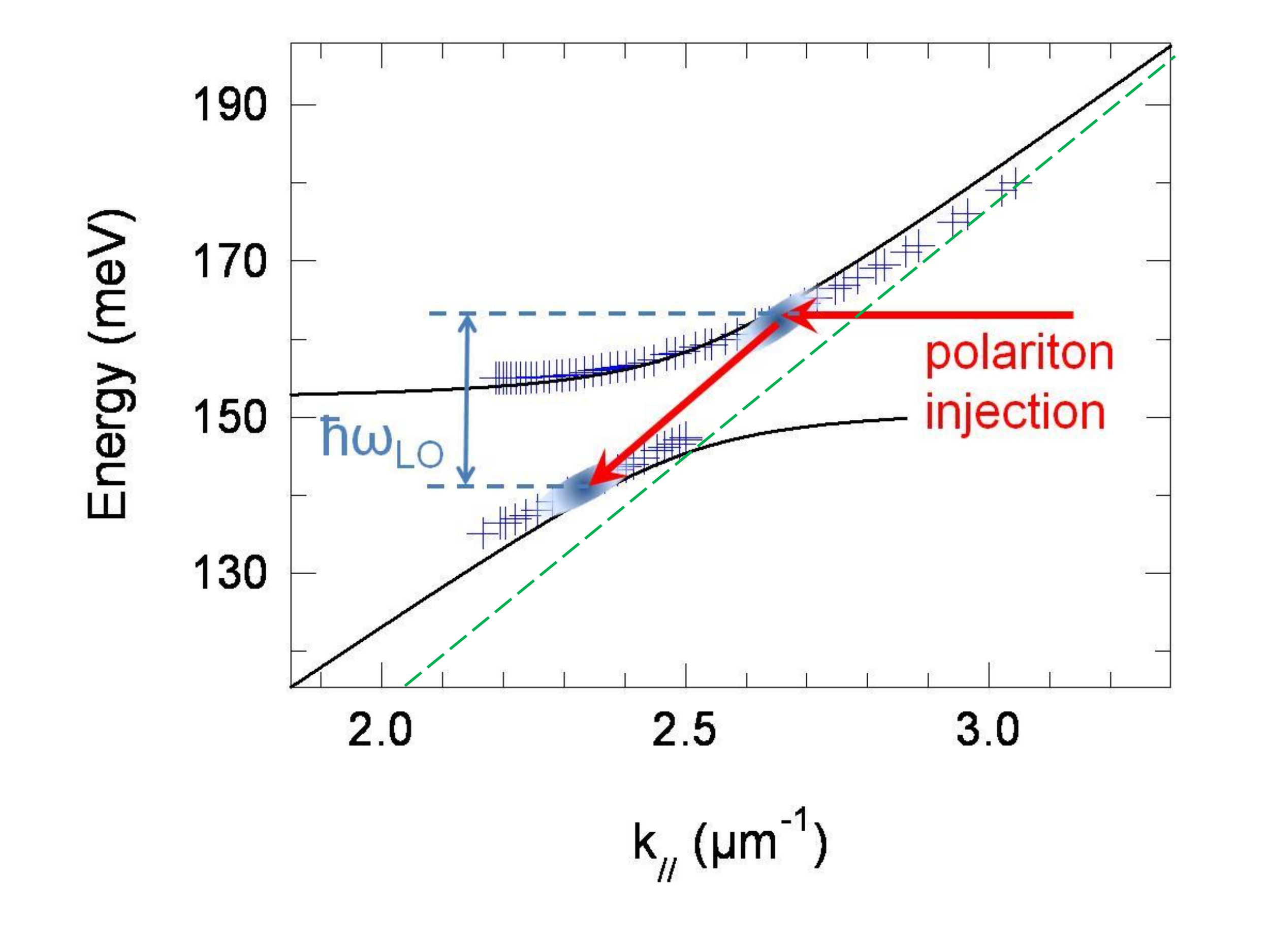}
\caption{(Colors online) EL maxima (crosses) measured at 77 K and 6~V, compared with the polaritonic dispersion obtained in absorption (continuous line). The dashed line indicates the light cone. We have also schematized the polariton population mechanism: electrical injection allows resonant pumping of the upper branch, while relaxation by LO-phonon emission allows the population of the lower branch.}
\label{schema}
\end{figure}

By changing the detection angle, we investigate the regime in which the intersubband transition and the cavity mode are in resonance. Angle resolved EL spectra measured at 6~V are shown in fig.~\ref{EL}c. The vertical lines give the energy position of the two bare transitions, as observed in panel (b). At low angle the spectra are very similar to those observed at $45 ^\circ$ because the system is far from the resonance condition. By increasing the internal angle of light propagation, we observe two peaks on both sides of $E_{21}$, whose energy position changes with the angle. We associate these peaks to EL features arising from the upper and lower polariton (respectively UP and LP) states. The third peak, at the energy $E_{32}$, is not angle dependent: it is associated to a weak coupling between the $3 \rightarrow 2$ transition and the second order cavity mode. It is important to notice that the two transitions in the main quantum well have a very different behavior as far as concern their coupling with the cavity mode. In fact the vacuum Rabi energy is proportional to $\sqrt{N_{g}-N_{e}}$, where $N_{g}$ (respectively $N_{e}$) is the electronic density of the ground (resp. excited) state of the transition. In our case only the fundamental subband is strongly populated, hence resulting in a strong coupling between the transition at $E_{21}$ and the cavity mode.
By comparing the spectra in panels b) and c) we can notice that the peaks associated to the upper polariton lie in the energy window corresponding to the transition between the injector and the state 2. This is analogous to what was observed in other electroluminescent polaritonic devices~\cite{sapienza_PRL, todorov_APL, jouy_PRB}, where the injector resonantly populates the lower polariton branch. In the present case, the high conduction band discontinuity of InAs/AlSb material system allows us to exploit the same mechanism to populate the polariton branches at higher energies.
It is important to stress that this injection mechanism cannot be invoked to explain the observation of the peak associated to the lower polariton in the spectra of fig.~\ref{EL}c. Indeed the quantum cascade structure does not present any transition at this energy. The observation of the lower polariton branch is attributed to the relaxation of polaritons from the upper to the lower branch. This mechanism is represented in fig.~\ref{schema}: the upper polariton branch is resonantly populated by the electronic injection, while the lower branch is populated thanks to the scattering between upper polaritons and LO-phonons. This mechanism has been theoretically studied~\cite{deliberato_PRL}, in presence of an optical pumping of the upper branch. It could lead to stimulated scattering of polaritons and to inversionless lasing. Figure~\ref{schema} also presents a comparison between the polaritonic dispersion extracted from the angle resolved EL spectra of fig.~\ref{EL}c and that resulting from absorption spectra (see ~\ref{struc}b). We can clearly see that the electroluminescent device allows us to observe the same polaritonic states as in absorption.

\begin{figure}[ht]
\centering
\includegraphics[width=0.8\columnwidth]{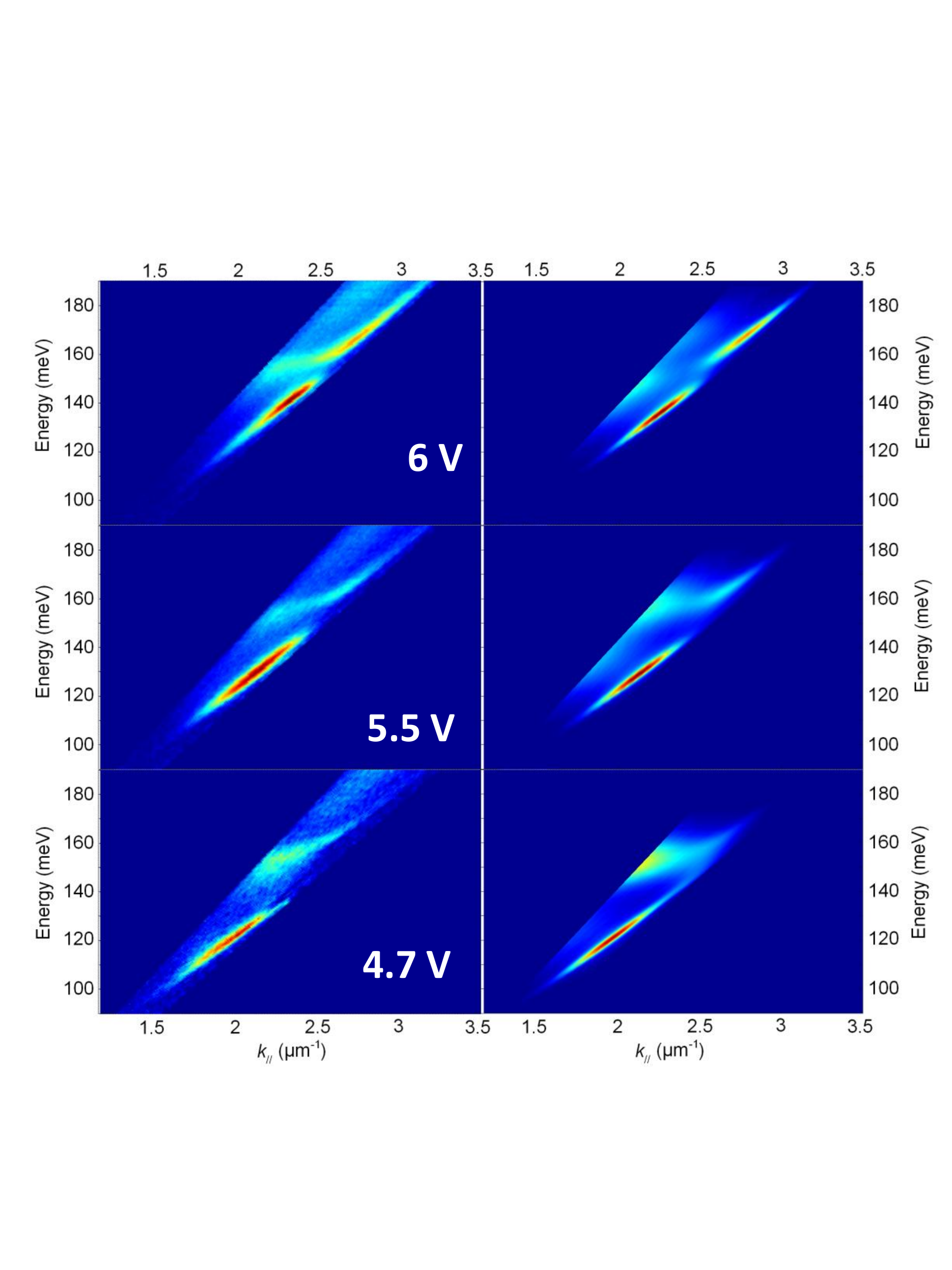}
\caption{(Colors online) EL (color scale) measured at 77 K as a function of the photon energy and in-plane wave vector. Left column presents the measured electroluminescence for 6~V (58 mA), 5.5~V (39 mA), 4.7~V (22 mA). Right column presents the simulated EL spectra for the corresponding voltages. We used the following parameters for the Gaussian functions: $\sigma=9$~meV for the three voltages, while $E_{inj}$ is equal to (from top to bottom): 168~meV, 160~meV, 151~meV.}
\label{kspace}
\end{figure}

The EL spectra at different voltages are presented as contour plots in figure~\ref{kspace} (left column). The spectra have been normalized by the function giving the ratio between the detected optical power and the emitted one, thus containing the contribution of the Fresnel coefficients~\cite{jouy_PRB}. In these graphs the EL intensity (color scale) is plotted as a function of the photon energy and in-plane wave vector ($k_{\sslash}$) for different applied voltages (from top to bottom respectively 6~V, 5.5~V, 4.7~V). At the three voltages we can clearly see the EL signal associated to the two polariton branches, anticrossing at the energy of the intersubband transition (which slightly Stark shifts with the voltage, from 154.5~meV at 6~V to 151.5~meV at 4.7~V). In the three bias conditions, the energy difference between the maxima of the electroluminescence is close to 30~meV, which is the energy of a LO-phonon in InAs. The comparison between the three panels shows that by changing the bias applied to the device, the energy windows, in which the electroluminescence associated to the polariton branches is observed, are modified. In fact the UP is resonantly populated by the injector at an energy depending of the bias~\cite{todorov_APL}, and the LP follows the same voltage dependence at an energy shifted by an LO-phonon. Similar features are observed in the EL spectra up to room temperature.

The EL spectra are simulated as $\mathcal{L}_{sim}\left(E,k_{\sslash}\right) = A \left(E,k_{\sslash}\right) \times n_{occ}(E)$, where $A \left(E,k_{\sslash}\right)$ is the absorption spectrum of the device, i.e. the optical density of states, while $n_{occ}(E)=n_{UP}\left(E \right) + n_{LP}\left(E \right)$ describes the occupation of the two polariton branches~\cite{stanley}.
 As the lower branch is populated by the relaxation of the upper polaritons, $n_{LP}$ is proportional to $n_{UP}$ via a coefficient depending on the upper and lower polariton energies, which are related to the voltage applied to the structure: $n_{LP}=C(V) \times n_{UP}$. This coefficient can also be expressed as the ratio between the radiative lifetime of the lower polariton (depending on the photonic Hopfield coefficient) and the polariton - LO-phonon scattering time, which depends on the matter part of the initial and final states. In our simulation, analogously to ref.~\onlinecite{sapienza_PRL,jouy_PRB}, we model the occupation of the upper polariton by a Gaussian function $n_{UP}\left(E \right)=\exp \left(-\frac{\left(E-E_{inj}\right)^2}{2\sigma^2}\right)$, where $E_{inj}$ is the injector energy and $\sigma$ its width.
$n_{LP}$ is described by an identical Gaussian function, but centered at the energy $E_{inj}-\hbar \omega_{LO}$.
The right column of fig.~\ref{kspace} shows the simulated EL spectra at the same voltages and in the same angular range as in the experiments. The simulated spectra reproduce very well all the experimental features: observation of the luminescence from both polariton branches and shift of the maxima of the electroluminescence with the voltage. The values $E_{inj}$ and $\sigma$ have been extracted from the low angle spectra~\cite{jouy_PRB}. For the three voltages, we found values of $C(V)$ of the order of unity (1.5, 2.3, 1.5 from the highest to the lowest voltage). This means that the polariton - LO-phonon scattering time is comparable with the lower polariton radiative lifetime, i.e. few hundreds of femtoseconds (0.1-0.3 ps). Furthermore, the lower polariton lifetime is limited by the radiative losses, and the intrabranch scattering is negligible.

\begin{figure}[ht]
\includegraphics[width=0.95\columnwidth]{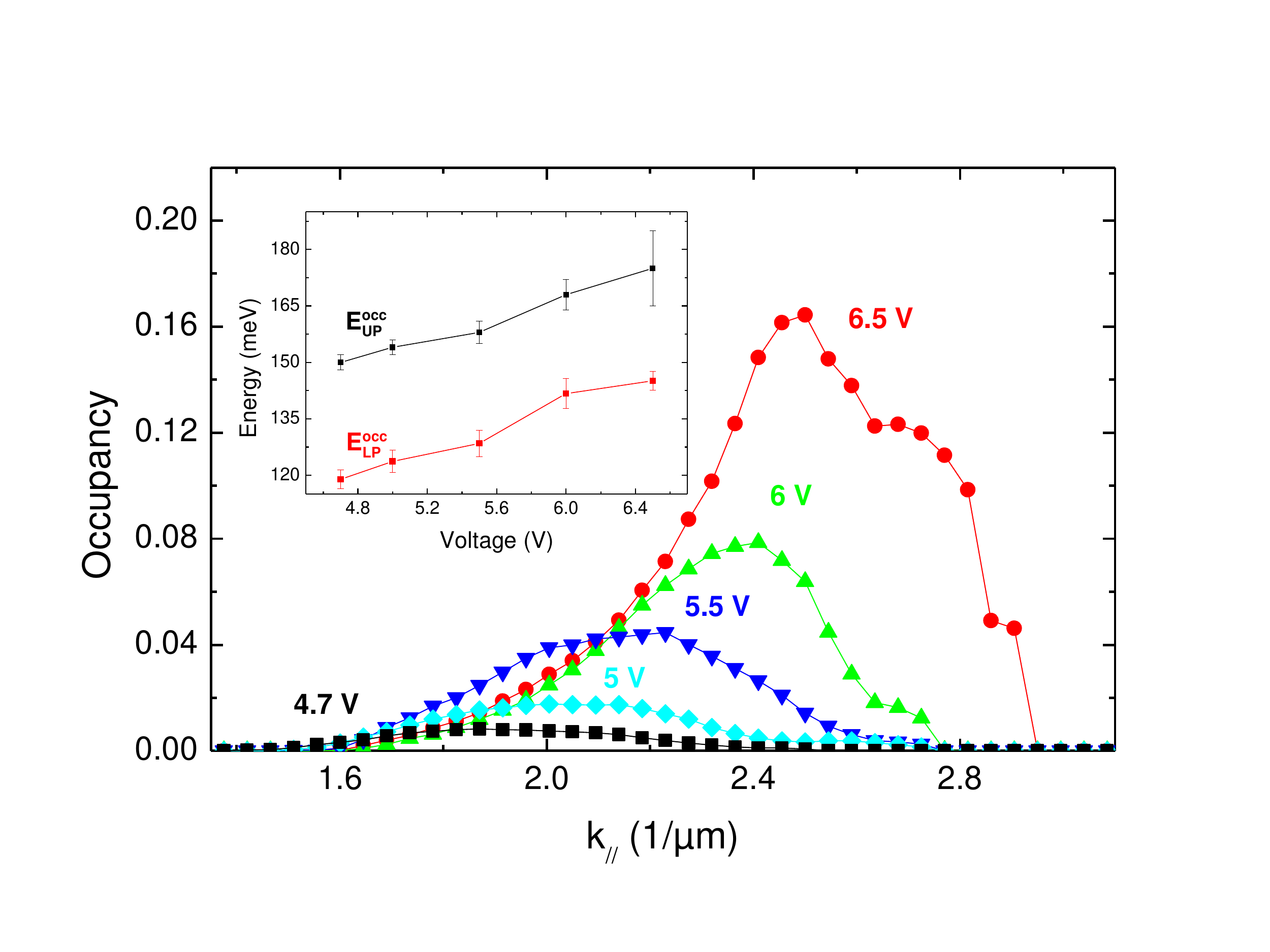}
\centering
\caption{(Colors online) Main panel: Lower polariton occupancy for different voltages. Inset: Energy evolution with the voltage of the LP and UP occupancy peaks. The mean value of the energy difference is $29.4 \pm 1.8$~meV, very close to the LO-phonon energy.}
\label{occup}
\end{figure}

The excellent agreement between simulated and measured spectra is a proof that the emission of an LO-phonon is the mechanism allowing the observation of the electroluminescence from the lower polariton branch. This interaction has been considered by De Liberato and Ciuti \cite{deliberato_PRL} as a possible mechanism for polariton stimulated emission. As a consequence, by following ref.~\onlinecite{stevenson}, we have estimated the occupation number for the lower polariton branch as a function of $k_{\sslash}$ from the measured electroluminescence.
We first estimate the power spectral density associated to the lower polariton branch after calibrating our HgCdTe detector. From this quantity, we deduce the LP density as a function of the in-plane wave vector, by assuming that the radiative lifetime of a polariton state is inversely proportional to its photon fraction, which is the square modulus of the Hopfield coefficient associated to the photonic part. We found a LP density of the order of $\approx 8.5 \times 10^5$~cm$^{-2}$ for the highest voltage. As the density of states per unit surface is given by $dn/dk_{\sslash}=k_{\sslash}/(2 \pi)$, we can deduce the occupancy of the LP branch, as a function of $k_{\sslash}$.
The results obtained for the different voltages are shown in figure~\ref{occup} (main panel). The peak value of the occupancy increases linearly with the current by a factor of 20 in the considered range, reaching the value of approximately~\cite{nota2} $0.16_{-0.03}^{+0.24}$ at 6.5 V (98 mA). The error that we estimate on this value is mainly related to the cavity photon lifetime, that we took equal to 0.2 ps from our simulated absorption spectrum. For higher voltages, we do not observe an increase in the occupancy as the phonon scattering populates the lower polaritons outside the cone of light. From figure~\ref{occup} (main panel), it is possible to see that the LP occupancy is with a good approximation a Gaussian function of $k_{\sslash}$ (up to 6 V), whose peak position shifts towards higher $k_{\sslash}$ while increasing the voltage. In the interval where the energy varies linearly with $k_{\sslash}$, this gives a Gaussian shape in energy for the LP occupancy, in agreement with the model presented above. This Gaussian function has the same width ($\approx 10$ meV) as the one associated to the injector. Note that, when increasing the voltage, the occupancy curves become asymmetric. This is due to the proximity of the light cone (indicated by a dashed line in figure~\ref{schema}), which cuts off the electroluminescence signal for high $k_{\sslash}$ values. The energies ($E_{LP}^{occ}$) corresponding to the peak occupancies at the different voltages are plotted in the inset of fig.~\ref{occup}, together with the maximum of the occupancy of the UP branch ($E_{UP}^{occ}$), which is given by the energy position of the injector. The two occupancy maxima shift towards higher energies by increasing the voltage. The energy difference between them is (within the error) 30~meV, which is the energy of a LO-phonon in InAs. This confirms that the occupancy of the LP is a phonon replica of that of the UP branch.

In conclusion, we provide evidence of intersubband polariton scattering with LO-phonons. The signature of this effect is in the electroluminescence spectra, where two peaks separated by the energy of an optical phonon appear for different injection condition. This process is realized by promoting electrons in the upper branch, at an energy which can be tuned with the voltage. Scattering with LO-phonons populates the lower polariton branch, thus opening a new radiative path. Occupancy of 0.16 is observed for the lower polariton states. An increase of this value toward one would lead to stimulated emission~\cite{stevenson} and laser without electronic population inversion. This could be readily obtained by decreasing the optical losses of the cavity and by modifying the polariton dispersion~\cite{todorov_PRL}.

\begin{acknowledgments}
The authors thank Y. Todorov and C. Ciuti for fruitful discussions. This work has been partially supported by the French National Agency (ANR) in the frame of its Nanotechnology and Nanosystems program P2N, project ANR-09-NANO-007. We acknowledge financial support from the ERC grant ``ADEQUATE''.
\end{acknowledgments}

\newpage

\end{document}